\documentstyle[manuscript,pra,aps]{revtex}
\begin{document}
\title{A novel realization of the Calogero-Moser scattering states as
coherent states}
\author{N. Gurappa\thanks{panisprs@uohyd.ernet.in}, P.S. Mohanty and
Prasanta K. Panigrahi\thanks{panisp@uohyd.ernet.in}} 
\address{School of Physics, 
University of Hyderabad,
Hyderabad,\\ Andhra Pradesh,
500 046 INDIA.}
\maketitle

\begin{abstract} 
A novel realization is provided for the scattering states of the
$N$-particle Calogero-Moser Hamiltonian. They are explicitly shown to be
the coherent states of the singular oscillators of the Calogero-Sutherland
model. Our algebraic treatment is straightforwardly extendable to a large
number of few and many-body interacting systems in one and higher
dimensions. 
\end{abstract}
\draft
\pacs{PACS: 03.65.-w, 03.65.Ge}

\newpage
The $N$-body Calogero-Sutherland model (CSM), containing harmonic and
inverse-square interactions, has been the subject of
intense investigation, since its inception \cite{cal,sut}. Its quantum
integrability, exact solvability and ability to capture the universal
aspects of diverse physical phenomena \cite{dpp}, has made this model
applicable to areas of physics, ranging from, fluid dynamics \cite{fd},
quantum Hall effect \cite{qhe}, to gauge theories \cite{gt}.

The many-particle Calogero-Moser model (CMM) \cite{cal,mos}, having only
pair-wise inverse-square interactions between particles, has also been
studied extensively in the context of novel statistics \cite{ns} and
recently, black hole physics \cite{bhp}. 

In this note, we give a novel realization of the scattering eigenstates
of this model as coherent states of the Calogero-Sutherland singular
oscillators. This, not only establishes a connection between the
wave functions of these two interacting many-body systems, but also
provides relationship between scattering and bound state problems. Our
algebraic procedure is straightforwardly extendable to a large class of
few and many-particle systems, both in one and higher dimensions.

The CSM Hamiltonian ($\hbar = m = \omega = 1$), 
\begin{equation}\label{bs}
H_{CSM} = - \frac{1}{2} \sum_{i=1}^N \frac{{\partial}^2}{\partial x_i^2}
+ \frac{1}{2} \sum_{1=1}^N x_i^2 + \frac {1}{2} g^2 \sum_{{i,j=1}\atop
{i\ne j}}^N \frac {1}{(x_i - x_j)^2}\qquad,
\end{equation}
describes bound-states, with energy spectrum given by $E_{\{n_i\}} = \sum_i
n_i + E_0$; here, only the ground-state energy $E_0 = N/2 + \alpha N (N -
1)/2$ depends on the coupling parameter with $\alpha = (1 + \sqrt{1 + 4
g^2})/2$. Apart from this shift, the rest of the spectrum, including the
degeneracy, is identical to that of $N$ decoupled harmonic oscillators.
Recently, a similarity transformation has been constructed which explicitly
maps the CSM to a set of decoupled oscillators \cite{ng}, hence naturally
explaining the above features of the model.

The CSM Hamiltonian without the harmonic term describes the Calogero-Moser
model:
\begin{equation}\label{ss}
H_{CMM} = - \frac{1}{2} \sum_{i=1}^N \frac{{\partial}^2}{\partial x_i^2}
 + \frac {1}{2} g^2 \sum_{{i,j=1}\atop
{i\ne j}}^N \frac {1}{(x_i - x_j)^2}\qquad.
\end{equation}
$H_{CMM}$ has only scattering states with energy eigenvalue $k^2/2$; this
is the same as that of $N$ free particles. Interestingly, in this
interacting system, there is no shift in the positions of the particles
due to scattering and the final configuration of the particle momenta
coincide with the initial ones up to a permutation. Furthermore, the
scattering phase shifts are energy independent.

In the following, we explicitly show that the scattering eigenfunctions of
the $N$-particle $H_{CMM}$ can be realized as a coherent state of the singular
oscillator of $H_{CSM}$, after removal of the Gaussian measure through a
similarity transformation. This is achieved by making use of (i) an
$SU(1,1)$ algebra \cite{cs-amp}, which contains the 
similarity transformed $H_{CSM}$ and $H_{CMM}$ as two of its generators and
(ii) a method, which enables one to find the canonical conjugate to a
given operator \cite{cc}.

It is easy to see that, the following generators 
\begin{eqnarray}
Z^{-1} (- H_{CMM}) Z &=& \frac{1}{2} \sum_{i=1}^N 
\frac{{\partial}^2}{\partial x_i^2} +  \alpha \sum_{{i,j=1}\atop{i\ne j}}^N
\frac {1}{(x_i - x_j)} \partial_i \equiv T_+ \quad,\nonumber\\  
\hat{S}^{-1} (- H_{CSM}/2) \hat{S}  &=& -\frac{1}{2} \left( \sum_i
x_i \partial_i + E_0 \right) \equiv T_0 \quad,\nonumber\\ 
\mbox{and}\qquad \qquad \qquad 
\frac{1}{2} \sum_i x_i^2 &\equiv& T_- \qquad,
\end{eqnarray}
satisfy the usual $SU(1,1)$ algebra:
$$
[ T_+ , T_- ] = - 2 T_0 \quad,\quad [ T_0 , T_{\pm} ] = \pm T_{\pm} \quad .
$$
Here, $Z \equiv \prod_{i<j} \mid(x_i - x_j)\mid^\alpha$ is the Jastrow factor
and $\hat{S} \equiv \exp\{- \frac{1}{2} \sum_i x_i^2\}\,\, Z\,\, \exp\{-
\frac{1}{2} T_+\}$. For the sake of convenience, we have chosen to deal
with $T_+$ instead of $H_{CMM}$, by removing the Jastrow factor. As
compared to $H_{CMM}$, $H_{CSM}$ has been subjected to an additional
similarity transformation which removes the Gaussian measure from the
wave functions. 

The quadratic Casimir, 
\begin{eqnarray}\label{qc}
\hat{C} = T_- T_+ - T_0 (T_0 + 1) = T_+ T_- - T_0 (T_0 - 1)\quad,
\end{eqnarray}
commutes with all the generators of the above $SU(1 , 1)$ algebra. The
unitary irreducible representations of this $SU(1,1)$ algebra, with $T_0$
diagonal, can be constructed in a straightforward manner. Since $T_0$
contains the Euler operator $\sum_i x_i \partial_i$, its eigenstates 
are homogeneous, symmetric polynomials of the particle coordinates;
the eigenvalue is given by $- (m + E_0)/2$, where $m$ is the degree of the
homogeneity. The eigenstates of $H_{CSM}$ can be obtained by an inverse
similarity transformation. One can immediately see that the
scattering states, modulo the Jastrow factor, {\it i.e.}, the eigenstates of
$T_+$, are the well-known coherent state of this $SU(1,1)$ algebra.
Analogous to the construction of the coherent states for the harmonic 
oscillator \cite{cs-amp}, the scattering states can be constructed
algebraically, starting from a basis where $T_0$ is diagonal. 

One needs the canonical conjugate of $T_+$ \cite{cc}:
\begin{equation}
[ T _+,  \tilde{T_-} ] = 1 \qquad .
\end{equation}
By choosing $\tilde{T_- } = T_-  F(T_0)$, one has
\begin{eqnarray}
[ T _+ ,  T_- F(T_0) ] = 
F(T_0) T_+ T_- - F(T_0 + 1 ) T_- T_+ = 1\quad. 
\end{eqnarray}

In terms of the quadratic Casimir, the above equation can be rewritten as
\begin{equation} \label{ft0}
F(T_0) \{ \hat C + T_0 (T_0  - 1 )\} - F ( T_0 + 1 ) \{ \hat C + T_0 ( T_0
+ 1)\} = 1 \qquad,
\end{equation} 
yielding,
\begin{equation}\label{ft01}
F(T_0 ) = \frac{-T_0 + a}{\hat C + T_0 (T_0 - 1)} \qquad.
\end{equation} 
The arbitrary constant $a$ and the
value of the quadratic Casimir $\hat C$ are to be fixed by demanding that
the above relations are valid in various sectors of the
Hilbert space. Recognizing that $T_+$ is the generalized Laplacian
operator \cite{cal}, the ground-state, $\phi_0(x)$, satisfying $T_+ \phi_0 = 0$, can be 
identified with the polynomials $P_m(x)$, whose existence and multiplicity 
have been proved by Calogero. Finding the explicit expressions for 
these homogeneous and permutation symmetric polynomials, for arbitrary $m$,
still remains an open problem \cite{sc}. After making use of the relations
$T_+ P_m(x) = 0$ and $T_0 P_m(x)= - [(m + E_0)/2] P_m$, Eq. (\ref{ft0}), when 
operated on the ground-state, yields $a = 1 - (E_0 + m)/2$. Similarly,
$$\hat{C} P_m(x) = (T_- T_+ - T_0 (T_0 + 1)) P_m(x) \quad,$$
gives $\hat{C} = \frac{1}{2}(m + E_0) (1 - (m + E_0)/2)$, for this unitary
irreducible representation. Using these values,
Eq. (\ref{ft01}) becomes
\begin{eqnarray}
F(T_0) P_m(x) &=& \frac{ - T_0 + a}{C + T_0 (T_0 - 1)} P_m(x) \nonumber\\
      &=& - \frac{1}{ T_0 - (m + E_0)/2} P_m(x) \qquad.
\end{eqnarray}

Much akin to the oscillator case, the eigenstates of $T_+ $ can be
obtained as a coherent state (unnormalized), starting from
\begin{equation}
T_+ P_m(x) = 0 \quad,
\end{equation}
and then performing a transformation by the displacement operator $U =
e^{\frac{1}{2} k^2 \tilde {T_-}}$:
\begin{equation}
U^{-1} T_+ U U^{-1} P_m(x) = 0\quad. 
\end{equation}  
It can be verified that
\begin{equation}
T_+ U^{-1} P_m(x) = - \frac{1}{2} k^2 U^{-1} P_m(x) \quad,
\end{equation}  
and the coherent state in the coordinate representation is
\begin{equation} \label{coher}
<x \mid k> = U^{-1} P_m(x) = e^{- \frac{1}{2} k^2 \tilde {T_-}} P_m(x)\quad.
\end{equation}  
Since, $T_+ P_m(x) = 0$, the above equation can also be rewritten as
\begin{eqnarray}
< x \mid k > &=& e^{- \frac{1}{2} k^2 \tilde{T_-}} e^{-T_+} P_m(x) \nonumber\\
&=& e^{ - \frac{1}{4} k^2 } e^{-T_+} e^{- \frac{1}{2} k^2 \tilde {T_-}}
P_m(x) \quad.
\end{eqnarray}
We have deliberately chosen these asymmetric factors, for the purpose of
comparing with earlier known results, as will be seen below. The explicit
form of the unnormalized, coherent state can be computed straightforwardly:
\begin{eqnarray}
< x \mid k > &=& e^{ - \frac{1}{4} k^2 } e^{-T_+} e^{- \frac{1}{2} k^2
\tilde {T_-}} P_m(x) \quad, \nonumber\\
&=& e^{- \frac{1}{4} k^2 } e^{-T_+} \sum_{n=0}^\infty \frac{(-
k^2/2)^n}{n!} (\tilde{T_-})^n P_m(x) \nonumber\\
&=& e^{ - \frac{1}{4} k^2} \sum_{n=0}^\infty \frac{(k^2/2)^n}{(E_0 + m
+ n)!} L_n^{(E_0 - 1 + m)}(r^2/2) P_m(x) \nonumber\\
&=& e^{\{ - k^2/4\}} (k/2)^{- (E_0 - 1 + m)}  (r)^{- (E_0 - 1 + m)}
J_{E_0 - 1 + m}(k r) P_m(x) \quad,
\end{eqnarray}
where, $r^2 = \sum_i x_i^2$ and $J_{E_0 - 1 + m}(k r)$
is the Bessel function.
In order to arrive at the above result, we made use of the
following result
$$T_+ \left(r^{2 n} P_m(x) \right) = 2 n (E_0 - 1 + m + n) r^{2
(n - 1)} P_m(x) \qquad,$$
and also the identity \cite{gra},
$$J_\alpha(2 \sqrt{x z}) e^z (xz)^{- \alpha/2} = \sum_{n=0}^\infty
\frac{z^n}{(n + \alpha + 1)!} L_n^\alpha(x)\quad.$$
After inserting the Jastrow factor one gets, 
\begin{eqnarray}
- \left(Z T_+ Z^{-1}\right) Z < x \mid k > &=& H_{CMM} Z < x \mid k >
\nonumber\\ &=&
\left(- \frac{1}{2} \sum_i \frac{\partial^2}{\partial x_i^2} + \frac{g^2}{2}
\sum_{i \ne j} \frac{1}{(x_i - x_j)^2}\right) Z < x \mid k > \nonumber \\ &=&
\frac{k^2}{2} Z < x \mid k > \qquad.
\end{eqnarray}

This result matches with the explicit solutions of Calogero \cite{cal},
once the center-of-mass degree freedom is removed. 

The above procedure can be extended to the $B_N$-type Calogero models
\cite{pr,ng1}, for which the Hamiltonian is given by,
\begin{equation} \label{bnbound}
H_{B_N} = - \frac{1}{2} \sum_{i=1}^N \partial_i^2
+ \frac {1}{2} \sum_{i=1}^N x_i^2 + \frac {1}{2} g^2
\sum_{{i,j=1}\atop {i\ne j}}^N \left\{\frac {1}{(x_i - x_j)^2} + 
\frac {1}{(x_i + x_j)^2}\right\} + \frac{1}{2} g_1^2
\sum_{i=1}^N \frac{1}{x_i^2} \qquad.
\end{equation}
The above Hamiltonian, without the harmonic term, {\it i.e.},
\begin{equation} \label{bnsca}
H_{\mbox{Sca.}} = H_{B_N} - \frac{1}{2} \sum_i x_i^2 \qquad,
\end{equation}
describes the scattering problem.
Akin to the CSM, one can define an $SU(1,1)$ algebra for this $B_N$ case:
\begin{eqnarray}
- Z^{- 1} H_{\mbox{Sca.}} Z &=& \frac{1}{2} \sum_i \partial_i^2 
+ \lambda \sum_{i<j} \frac {1}{(x_i^2 - x_j^2)} (x_i \partial_i - x_j
\partial_j)  + \lambda_1 \sum_i \frac{1}{x_i} \partial_i \equiv K_+
\quad,\nonumber\\   
\hat{T}^{-1} (- H_{B_N}/2) \hat{T} &=& -\frac{1}{2} \left( \sum_i
x_i \partial_i + \epsilon_0 \right)
\equiv K_0 \quad,\nonumber\\  
\mbox{and}\qquad \qquad \qquad 
\frac{1}{2} \sum_i x_i^2 &\equiv& K_- \qquad,
\end{eqnarray}
where, $g^2 = \lambda (\lambda - 1)$, $g_1^2 = \lambda_1 (\lambda_1 - 1)$
and $\epsilon_0 = N (\frac{1}{2} + (N-1) \lambda + \lambda_1)$ is the
ground-state energy. Here, $\hat{T} \equiv Z  \exp\{- \frac{1}{2}\sum_i x_i^2\}
\exp\{- K_+/2\}$ and $Z \equiv \prod_{1\le {j < {k\le N}}}|x_i - x_j|^\lambda
|x_i + x_j|^\lambda \prod_k^N |x_k|^{\lambda_1}$. Knowing the homogeneous
and symmetric polynomials, $Q_m(x)$: $K_+ Q_m(x) = 0$, one can
easily construct the scattering solutions of Eq. (\ref{bnsca}) as a
coherent state of the $B_N$-type Calogero model, as has been done for the
CSM case. The same technique can be applied to other one and higher
dimensional models \cite{oh1,oh2,oh3,oh4,oh5}.

In conclusion, although the solutions of the bound and scattering state
problems were independently known earlier, the connection we have
established between them is novel. Here, we would like to point out that
the coherent states for the two-particle CSM have been found by Agarwal
and Chaturvedi by explicitly solving the corresponding differential
equations \cite{as}. The present method is, not only, purely algebraic
and applicable to an arbitrary number of particles, but also, makes it
explicit that, these coherent states, after removal of the Gaussian
measure, are nothing but the scattering states of the Calogero-Moser
Hamiltonian. It will be of great interest to apply the method developed
here to other interacting systems.

N.G thanks U.G.C (India) for the financial support through the S.R.F
scheme.

\end{document}